\documentclass[aps,prb,superscriptaddress,showkeys,twocolumn]{revtex4-1}
\usepackage{amsmath,amssymb}
\usepackage{hhline}
\usepackage{graphicx}% Include figure files
\usepackage{dcolumn}% Align table columns on decimal point
\usepackage{bm}% bold math
\usepackage[latin1]{inputenc}
\usepackage{float}
\usepackage{color}
\usepackage{amsmath}
\definecolor{rojo}{rgb}{1,0,0}
\definecolor{verde}{rgb}{0,0.8,0.5}
\definecolor{azul}{rgb}{0,0,1}
\definecolor{rosa}{cmyk}{0,1,0,0}
\usepackage[toc,page]{appendix} 
\usepackage{latexsym}
\usepackage{amsmath}
\usepackage{mathrsfs}
\usepackage{bm}
\usepackage{amsthm}
\usepackage{physics}
\usepackage{adjustbox}
\usepackage{bbm}
\usepackage{amsfonts}
\usepackage{amssymb}
\usepackage{color}
\usepackage{epsfig}
\usepackage{hyperref}
\hypersetup{colorlinks=true, citecolor=blue}
\usepackage{soul}
\usepackage[normalem]{ulem}
\usepackage[cmtip,all]{xy} 
\usepackage{comment}
\usepackage{multirow}
\usepackage{mathtools}
\newcolumntype{L}{>{$}l<{$}}

\newcommand{\longsquiggly}{\xymatrix{{}\ar@{~>}[r]&{}}} 

\begin{document}

\title{Mechanically modulated spin orbit couplings in oligopeptides}

\author{Juan Daniel Torres}
\affiliation{Yachay Tech University, School of Physical Sciences \& Nanotechnology, 100119-Urcuqu\'i, Ecuador}
\author{Raul Hidalgo}
\affiliation{Yachay Tech University, School of Physical Sciences \& Nanotechnology, 100119-Urcuqu\'i, Ecuador}

\author{Solmar Varela}
\email{svarela@yachaytech.edu.ec}
\affiliation{Yachay Tech University, School of Chemical Sciences \& Engineering, 100119-Urcuqu\'i, Ecuador}

\author{Ernesto Medina}
\email{emedina@yachaytech.edu.ec}
\affiliation{Yachay Tech University, School of Physical Sciences \& Nanotechnology, 100119-Urcuqu\'i, Ecuador}
\affiliation{Simon A. Levin Mathematical, Computational and Modeling Sciences Center, Arizona State University, P.O. Box 873901, Tempe, AZ 85287-3901, United States}%

\date{\today}

\begin{abstract}
Recently experiments have shown very significant spin activity in biological molecules such as DNA, proteins, oligopeptides and aminoacids. Such molecules have in common their chiral structure, time reversal symmetry and the absence of magnetic exchange interactions. The spin activity is then assumed to be due to either the pure Spin-orbit (SO) interaction or SO coupled to the presence of strong local sources of electric fields. Here we derive an analytical tight-binding Hamiltonian model for Oligopeptides that contemplates both intrinsic SO and Rashba interaction induced by hydrogen bond. We use a lowest order perturbation theory band folding scheme and derive the reciprocal space intrinsic and Rashba type Hamiltonian terms to evaluate the spin activity of the oligopeptide and its dependence of molecule uniaxial deformations. SO strengths in the tens of meV are found and explicit spin active deformation potentials. We find a rich interplay between responses to deformations both to enhance and diminish SO strength that allow for experimental testing of the orbital model. Qualitative consistency with recent experiments shows the role of hydrogen bonding in spin activity.

\end{abstract}

\maketitle

%\tableofcontents
%****************************************************************************************************************************
\section{Introduction}
There has been considerable interest recently in the electron spin polarizing ability of biological chiral molecules such as DNA, proteins, oligopeptides and aminoacids\cite{NaamanDNA,Gohler,NaamanPhotosystem,Aragones,NaamanAminoacids}. The effect known as Chiral-Induced Spin Selectivity (CISS) is impressive since the electron polarizations achieved, both for self assembled monolayers and single molecule setups, exceeds those of ferromagnets\cite{PaltielMemory}. The qualitative explanation for spin activity in the absence of a time reversal symmetry breaking interaction has been suggested to be due to the atomic spin-orbit coupling\cite{Sina,MedinaLopez}. Although the small size of the interaction has suggested invoking sources such as inelastic effects\cite{GuoSun,Bart,VarelaInelastic}, recent works have shown that tunneling alone can exponentiate the small spin-orbit values to yield very high polarizations\cite{TunnelingSO}.

Analytical tight-binding modelling has proven very powerful to understand the qualitatively new features of low dimensional systems. An emblematic examples is the discovery of topological insulators \cite{KaneMele} and the integer quantum hall effects without magnetic fields\cite{Haldane}. In the context of the CISS effect, a recent model\cite{Varela2016} described the spin activity of DNA on the basis of a tight-binding (TB) model that assumes mobile electrons on the $\pi$ orbitals of the bases and the spin-orbit coupling (SOC) due to the intra-atomic interactions of C, O, N. The resulting model yields a consistent picture of how a time reversal symmetric Hamiltonian can result in spin-polarization. A more recent analytical TB model have also described transport features of Helicene\cite{HeliceneMujica}.

While attempting to assess the dominant player in electron spin transport on large molecules, an opportunity arises to validate the orbital model using mechanical deformations\cite{Kiran}. The spin polarization response hints at the orbital participation involved in determining the SOC strength\cite{Varela2018,VarelaJCP}. One can then also perform transport and determine the behaviour of a finite system with coupling to reservoir details.

In this work, we derive an analytical tight-binding Hamiltonian model for oligopeptides that assumes that the basic ingredients are: i) the atomic SO interaction from double bonded (orbital) oxygen atoms in the amine units provide the transport, ii) the Stark interaction matrix element between the $p_z$ orbital and the oxygen $s$ orbital produced by the hydrogen bond polarization, and iii) overlaps between nearest neighbor oxygen orbitals.  We use perturbation theory band folding scheme and derive the real space and reciprocal space Rashba and intrinsic SOC Hamiltonian to evaluate the spin activity of the oligopeptide. 

The paper is organized as follows: in Sec. \ref{section2} we first introduce the full tight binding model of the oligopeptide including both the Stark and the SOC. Then we use band folding to reduce an $8\times8$ space encompasing the orbital space to a $2\times2$ effective space involving one effective $p_z$ per site. Thus we derive the resulting Rashba and intrinsic SOC's and energy corrections. We find closed form expressions for dependencies of the interactions on the geometry of the molecule and the type of amino-acid units. There arise four different SOC terms: two associated to the Rashba interaction and two to the intrinsic coupling. In Section \ref{sectionIII} we obtain the Hamiltonian in reciprocal space by way of a Bloch expansion. In Sec. \ref{sectionIV} we show the analysis of the behavior of the SOC magnitudes under deformations. The interplay between these spin active interactions yield opposite responses to the longitudinal mechanical deformations, with predominance of the SO enhanced stretching. Furthermore, the Rashba coupling, depending on the polarization of the hydrogen bond, yields additional enhanced SO due to stretching as reported experimentally for oligopeptides\cite{Kiran}. These results point to role of the atomic SO and hydrogen bonding in the spin activity of biological molecules. Finally, in Section \ref{conclusions} we offer a Summary and Conclusions.

%****************************************************************************************************************************

\section{Tight Binding Model}\label{section2}

\begin{figure}[h!]
	  \centering
	  \includegraphics[width=6cm]{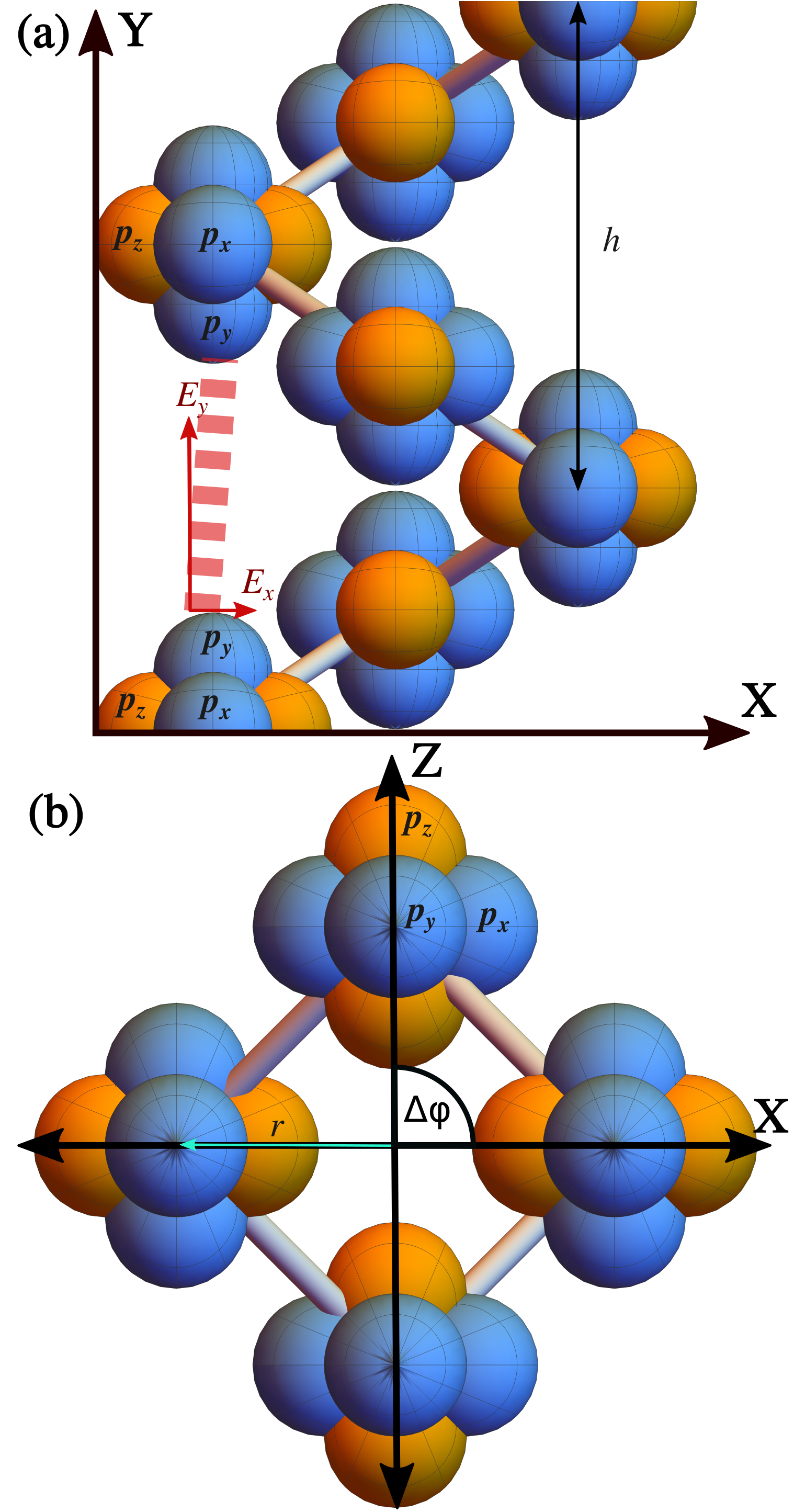}
	  \caption{(a) Front view of the helical oligopeptide in the $XY$-plane. The pitch of the helix is indicated as $h$ and labels for each $p$orbital are shown. The internal electric field caused by the hydrogen bond and the component along each direction are shown in red. (b) Top view of the helical oligopeptide in the $XZ$-plane where $r$ represents the radius of the helix, and $\Delta \varphi$ is the angle between consecutive bases.}
\label{fig:structure}
\end{figure}

Consider a helix as shown in Fig.\ref{fig:structure}. Each atom is described by a set of $\{ s,p_x,p_y,p_z \}$ orbitals associated with valence oligopetide constituents such as C, N, O. The mobile electrons are assumed to be provided with the double bonded oxygen (carboxyl group) attached by hydrogen bonding (see Fig.\ref{fig:structure}a) to the amine group in the oligopeptide. The backbone of the molecule is bonded through the $\{ s,p_x,p_y \}$ orbitals that lie tangentially to structure. We consider that the electrons associated to these bonds do not contribute to transport. The $\pi$ structure of the double-bonded oxygen is accounted for by the remaining $p_z$ orbitals in the radial direction (see Fig.\ref{fig:structure}b) akin to the structure of a single walled nanotube. 
%The $p$-orbital representation in terms of spherical harmonics $| l m \rangle$ %is given by
%\begin{equation}
%\begin{split}
%	\ket{p_x} &= - \frac{1}{\sqrt{2}} (\ket{1,1} - \ket{1,-1}),\\
%	\ket{p_y} &= \frac{i}{\sqrt{2}} (\ket{1,1} + \ket{1,-1}),\\
%	\ket{p_x} &= \ket{1,0}.
%\end{split}
%\end{equation}

The axis of the chain is considered along the Y-axis with a set of orbitals on sites $\imath$, such that $\imath=1,...,N$.  The position $R_{\imath}$ in fixed or global coordinate system (XYZ) can be written as 
\begin{multline}
	\mathbf{R_{\imath}} = r \cos[(\imath-1)\Delta\varphi] \mathbf{e_Z}\\ + r \sin[(\imath-1)\Delta\varphi] \mathbf{e_X} + h\frac{(\imath-1)\Delta\varphi}{2\pi}  \mathbf{e_Y},
\end{multline}
where $r$ is the radius of the helix, $h$ is the pitch, and $\Delta\varphi$ represents the angle between the positions of two consecutive sites. The vector that connects two sites $\imath$ and $\jmath$ of the helix is $\mathbf{R_{\jmath \imath}} = \mathbf{R_{\jmath}} - \mathbf{R_{\imath}}$. 

Electrons are well coupled along the helical structure (as opposed to the coupling from one turn of the helix to the next) and different couplings are included. The full Hamiltonian of the system can be written in the form 
\begin{equation}\label{H}
    H = H_{K} + H_{SO} + H_{S},
\end{equation}
where $H_{K}$ is the kinetic term or the bare Slater-Koster overlaps, $H_{SO}$ include the Spin-Orbit (SO) interactions, and $H_{S}$ is the Stark interaction resulting from electric dipoles (hydrogen bonding) in the molecule. 

\subsection{Stark interaction and hydrogen bonding}
In a helical peptide, the hydrogen bonds between the amino and carboxyl groups stabilize the helical structure\cite{HydrogenBond}. As shown in ref.\cite{Blanco-Ruiz,Varela2018}, the near field electrostatics of the bond yield among the highest electric field one finds in a molecules that goes unscreened. These electrostatic fields, have been proposed to generate local interactions that open new transport channels. In the model, the Stark interaction associated with hydrogen bond polarization couples $s$ with $p$ orbitals on the double bonded oxygen of the carboxyl group along the direction of the dipole field in the form $H_S = -e \mathbf{E} \cdot \mathbf{r}$ where $\mathbf{E}$ is the electric field (see ref.\cite{Varela2018}), $\mathbf{r}$ is the position vector of the atom and $e$ is the electron charge. In spherical, local, coordinates we have
\begin{equation}\label{Stark}
	 H_S = - e r (E_x \sin\theta \cos\varphi + E_y \sin\theta \sin\varphi),
\end{equation}
where $E_{x,y}$ represents the components of the electric field in the indicated directions (red arrows in Fig.\ref{fig:structure}). The source of the electric field along $x$ and $y$ was obtained from ref.[\onlinecite{Varela2018}] where the electric field was computed accounting for the local dipole field of hydrogen bonding. 

%To compute the Stark matrix elements we consider the real space representation of the hydrogenic atomic orbitals in spherical coordinates.
%\begin{equation}
%\begin{split}
%    s(r,\theta,\phi) &= \langle \mathbf{r}|s\rangle = \frac{Z^3}{\sqrt{8\pi %a_o^3}} e^{-rZ/2a_o} \left( 1-\frac{Zr}{2a_o} \right),\\
%    p_x(r,\theta,\phi) &= \langle \mathbf{r}|p_x\rangle = %\frac{Z^3}{\sqrt{32\pi a_o^3}} e^{-rZ/2a_o} \sin\theta \cos\phi,\\
%    p_y(r,\theta,\phi) &= \langle \mathbf{r}|p_y\rangle = %\frac{Z^3}{\sqrt{32\pi a_o^3}} e^{-rZ/2a_o} \sin\theta \sin\phi,
%\end{split}
%\end{equation}
%where $a_o$ is the Bohr radius. 

In general, the hydrogen bond direction has a component both along the $x$ and $y$ directions. However, the component along $x$ direction is much smaller than the $y$ component, since the bond is essentially in the $Y=y$ direction. Then, consider $\xi_{sx}$ and $\xi_{sy}$, these are given by,
\begin{equation}
\begin{split}
   \xi_{sx}  = \langle s | H_S| p_x\rangle, \qquad \xi_{sy}  = \langle s | H_S| p_y\rangle.
   \end{split}
\end{equation}
In the case of mechanical deformation, higher order terms may be relevant when the helix is stretched. 

\subsection{Spin-Orbit interactions}
The SO interaction has been well described by tight-binding treatments in the context of low dimensional systems\cite{Huertas2006,Konschuh2010,Varela2016}. The atomic SO interaction couples the spin of the electron to the internal electric field of the nuclei. The SO Hamiltonian is
\begin{equation}
\begin{split}
		H_{SO} &=\frac{e}{2m^2_oc^2}(\nabla V \times \textbf{p})\cdot\textbf{S},\\
	%	&=\frac{1}{r}\frac{\partial V}{\partial r} 
     %   \frac{e}{2m^2_oc^2}\textbf{L}\cdot\textbf{S},\\
	   &=\Gamma\textbf{L}\cdot\textbf{S}, \\
	%	&=\frac{\lambda}{2}(L_+S_- + L_-S_+ +2L_zS_z),
\end{split}
\end{equation}
where $V$ is electrical potential of the nuclei as seen by valence electrons of the orbital basis, $m_o$ is the rest electron mass, $e$ is the charge of the electron, $c$ is the speed of light, $\textbf{S}$ and $\textbf{L}$ are the spin and orbital angular momentum operators, respectively. The SO matrix elements couple the basis $p$ orbitals as shown in Table \ref{tab:SO},
\begin{table}[!htbp]
\caption{SO matrix elements between $p$ orbitals in the local coordinate system.}
\begin{center}
\begin{tabular}{| L | L | L | L | L |}
\hline
 & |p_x\rangle & |p_y\rangle & |p_z\rangle \\
\hline
\langle p_x|& 0 & -i \xi_p \mathbf{s_z} & i \xi_p \mathbf{s_y}\\
\langle p_y|&i \xi_p \mathbf{s_z}&0&-i \xi_p \mathbf{s_x}\\
\langle p_z|&-i \xi_p \mathbf{s_y}&i \xi_p \mathbf{s_x}&0\\
\hline
\end{tabular}
\end{center}

\label{tab:SO}
\end{table}
where $\xi_p=\Gamma/2$ is the magnitude of the SO interaction for $p$ orbitals and $s_{j}$ are the Pauli matrices in the  rotating coordinate system. The rotated spin operators, i.e. the spin operators in the local frame, are
\begin{equation}
\begin{split}
    \mathbf{s}_x &= -\sin(\varphi_i) \sigma_x+\cos(\varphi_i)\sigma_z,\\
    \mathbf{s}_y&=\sigma_y,\\
    \mathbf{s}_z&=\cos(\varphi_i) \sigma_x+\sin(\varphi_i)\sigma_z.
\end{split}
\end{equation}
There are two relevant SO interactions that lead to different spin active processes. The first is the intrinsic SO interaction, which is the pure matrix element between atomic orbitals, i.e. $H_{SO}$. The paths of the first order intrinsic SO are,
\begin{equation}\label{Intrinsic}
    p_z^\imath \rightarrow E_{zx}^{\imath \jmath} \rightarrow p_{x}^\jmath \rightarrow \xi_p \rightarrow p_z^\jmath,
\end{equation}
\begin{equation}\label{Intrinsic}
    p_z^\imath \rightarrow E_{zy}^{\imath \jmath} \rightarrow p_{y}^\jmath \rightarrow \xi_p \rightarrow p_z^\jmath,
\end{equation}
where the Slater-Koster (SK) overlaps $E_{\mu\mu'}^{\imath\jmath}$ between an orbital $\mu$ on $\imath$ site and orbital $\mu'$ on site $\jmath$, are defined in the appendix \ref{sec:SK}. The second type of SO interaction is possible when there is Stark interaction. The Rashba SO interaction arises as a combination of both the Stark interaction and the bare SOC. The Stark interaction has been argued to be the strongest source of electric fields in molecules outside the vicinity of the nucleus\cite{Varela2018} because of the presence of hydrogen bond polarization in the near field\cite{Blanco-Ruiz}. The paths of a first order Rashba process are
\begin{equation}\label{Rashba1}
    p_z^\imath \rightarrow E_{zs}^{\imath \jmath} \rightarrow p_s^\jmath \rightarrow \xi_{sx} \rightarrow p_{x}^\jmath \rightarrow \xi_p \rightarrow p_z^\jmath,
\end{equation}
\begin{equation}\label{Rashba2}
    p_z^\imath \rightarrow E_{zs}^{\imath \jmath} \rightarrow p_s^\jmath \rightarrow \xi_{sy} \rightarrow p_{y}^\jmath \rightarrow \xi_p \rightarrow p_z^\jmath.
\end{equation}
Geometrical details of the problem determine the effective SO magnitudes resulting from the interplay between different first order transport processes, e.g. interference between (\ref{Rashba1}) and (\ref{Rashba2}).

\subsection{Effective Hamiltonian}
The Hamiltonian of Eq.(\ref{H}) in the basis of atomic orbitals can be written as
\begin{equation}\label{Hfull}
    H = \left( \begin{array}{cc}
        H_{\pi} & T \\
        T^\dagger & H_{\sigma}
    \end{array} \right), 
\end{equation}
where $H_{\pi}$ and $H_{\sigma}$ are the structural Hamiltonians and $T$ correspond to the connection between $\pi$ and $\sigma$ spaces.
In Table \ref{tab:Hfull}, all the matrix elements of the full Hamiltonian are written explicitly. Here, the SK overlaps are represented by $V_s$, $V_x$, $V_y$, and $V_z$ (see appendix A), $\epsilon_{p}^{\sigma}$ is the site energy for the bonded orbitals $p_x$ and $p_y$, $\epsilon_{p}^{\pi}$ is the site energy of the orbital $p_z$, and $\epsilon_s$ is the energy of the orbital $s$. 

\begin{table}[!htbp]
\caption{The matrix elements of the full Hamiltonian in the local coordinate system. The $\pi$ and $\sigma$ spaces are the diagonal components while the off diagonal correspond to $T$ and $T^\dagger$ of (\ref{Hfull}). }
\begin{center}
\adjustbox{max width=\textwidth}{
\begin{tabular}{| L | L  L | L  L  L  L  L  L |}
\hline
 & |p_z\rangle_i & |p_z\rangle_j &  |s\rangle_i & |p_x\rangle_i & |p_y\rangle_i & |s\rangle_j & |p_x\rangle_j & |p_y\rangle_j \\
\hline
\langle p_z|_i & \epsilon_p^\pi & V_z & 0 & -i\xi_{p}\mathbf{s_y} & i\xi_{p}\mathbf{s_x} & V_s & V_x & V_y \\
\langle p_z|_j & V_z & \epsilon_p^\pi & V_s & -V_x & -V_y & 0 & -i\xi_{p}\mathbf{s_y} & i\xi_{p}\mathbf{s_x}\\ 
\hline
\langle s|_i & 0 & V_s & \epsilon_s & \xi_{sx} & \xi_{sy} & 0 & 0 & 0 \\
\langle p_x|_i & i\xi_{p}\mathbf{s_y} & -V_x & \xi_{sx} & \epsilon_p^\sigma & 0 & 0 & 0 &0 \\
\langle p_y|_i & -i\xi_{p}\mathbf{s_x} & -V_y & \xi_{sy} & 0 & \epsilon_p^\sigma &  0 & 0 & 0 \\
\langle s|_j & V_s & 0 & 0 & 0 & 0 & \epsilon_s & \xi_{sx} & \xi_{sy} \\
\langle p_x|_j & V_x & i\xi_{p}\mathbf{s_y} & 0 & 0 & 0 & \xi_{sx} & \epsilon_p^\sigma & 0 \\
\langle p_y|_j & V_y & - i\xi_{p}\mathbf{s_x} & 0 & 0 & 0 & \xi_{sy} & 0 & \epsilon_p^\sigma\\
\hline
\end{tabular}}
\end{center}
\label{tab:Hfull}
\end{table}

The goal is to obtain an effective Hamiltonian that describes the $\pi$-space including the physics of the $\sigma$-space as a perturbation. For this purpose, we use an energy independent perturbative partitioning approach developed by L\"owdin \cite{Lowdin1,Lowdin2,Lowdin3,boykin}. The Band Folding (BF) method is used to obtain an effective Hamiltonian using matrix perturbation theory. It is a transformation in the same sense of the Foldy-Wouthuysen transformation\cite{FW} maintaining only first order corrections. The effective Hamiltonian for the $\pi$-structure is
\begin{equation}
    \mathcal{H}\approx H_{\pi} - T H_{\sigma}^{-1} T^\dagger.
\end{equation}
 No additional corrections arise from wavefunction normalization\cite{MccannBilayer}. Then, one simplifies the problem from  $8\times8$, in orbital and site space, to $2\times 2$. Spin active terms are written implicit. The effective Hamiltonian is,
\begin{equation}
    \mathcal{H}=\left( \begin{array}{cc}
        \epsilon_\pi  & V_z - i ((\bm{\alpha} + \bm{\lambda})\times \mathbf{s})_z\\
         V_z + i ((\bm{\alpha} + \bm{\lambda}) \times \mathbf{s})_z & \epsilon_\pi
    \end{array} \right).
    \label{Heffective}
\end{equation}
%In the effective Hamiltonian (\ref{Heffective}), the energies are shifted due to kinetic, intrinsic SO, and Stark interactions of higher orders as,
%\begin{equation}
%\begin{split}
%    \epsilon_\pi &= \epsilon_p^\pi 
%    - \left(\frac{\xi _p^2}{\epsilon _{\text{px}}}+\frac{\xi %_p^2+V_y^2}{\epsilon_{\text{py}}}+\frac{V_x^2}{\epsilon_{\text{px}}}+\frac{V_s^%2}{\epsilon
%   _s}\right)\\&
%   +\frac{\xi _{\text{sy}}^2 \left(\epsilon _s \epsilon _{\text{px}} %V_y^2+\epsilon_{\text{px}} \epsilon _{\text{py}} V_s^2\right)}{\epsilon %_{\text{px}} \epsilon_{\text{py}}^2 \epsilon _s^2}.
%    \end{split}
%\end{equation}
%Also, there are diagonal shifts coming from the Stark interaction that break sub-lattice symmetry, but their contribution arise from processes inside the $\sigma$ space that do not contribute. The energy difference, $\epsilon_S$, is given by,
%\begin{equation}\label{Sshift}
 %   \epsilon_S = \frac{ V_s \xi _{\text{sy}} V_y}{\epsilon _{\text{py}} \epsilon _s}+\frac{ \epsilon _{\text{py}} V_s \xi _{\text{sx}}
  
% V_x}{\epsilon _{\text{px}} \left(\epsilon _{\text{py}} \epsilon _s-\xi
 %  _{\text{sy}}^2\right)}.
%\end{equation}

%{\color{red} In this section we should  include explicitly the terms associate with the specific interactions, not only the corrections. In this way, we can determmine the magnitude of the interactions. }

There are intrinsic SO linear in $\xi_{p}$ and Rashba bilinear in $\xi_{p}\xi_{sy}$ interactions that contribute to the total SO interaction. The intrinsic SOC contribution between sites $\imath$ and $\jmath$ is given by,
\begin{equation}\label{IntrinsicSO}
    \mathcal{H}_{so}^{\imath \jmath} = i (\alpha_x s_y - \alpha_y s_x) = i (\bm{\alpha}\times\mathbf{s})_z,
\end{equation}
 where $\mathbf{s}$ is the vector of Pauli matrices and $\bm{\alpha}$ is the vector with the magnitude of the intrinsic SO in each coordinate that are defined as,
 \begin{equation}
    \alpha_x=\frac{2 \xi_p V_x}{\epsilon_p}, \qquad  \alpha_y=\frac{2 \xi_p V_y}{\epsilon_p}.
 \end{equation}
 The estimated values, considering characteristic values for the oligopeptide, are $\alpha_x\sim8.97$ meV and $\alpha_y\sim10.20$ meV (see appendix B).
 The Rashba SO has contributions from higher order terms from the Stark interaction in the form,
 \begin{equation}\label{RashbaSO}
    \mathcal{H}_{R}^{\imath\jmath} = i (\lambda_x s_y - \lambda_y s_x) =  i(\bm{\lambda}\times\mathbf{s})_z,
\end{equation}
 where $\bm{\lambda}$ is a vector with the Rashba SO magnitude in each component. They are given by,
\begin{equation}
\begin{split}
    \lambda_x &= \frac{\xi_p (\xi_{\text{sy},\imath}-\xi_{\text{sy},\jmath})V_s}{\epsilon_{\text{pz}}\epsilon_{\text{s}}}-\frac{2 \xi _p \epsilon _{\text{py}}^2 \epsilon _s \xi _{\text{sx}}^2 V_x}{\epsilon_{\text{px}}^2 \left(\xi _{\text{sy}}^2-\epsilon _{\text{py}} \epsilon
   _s\right){}^2}  \\
   &+\frac{2 \xi _p \xi _{\text{sx}} \xi _{\text{sy}} V_y}{\epsilon_{\text{px}} \left(\xi _{\text{sy}}^2-\epsilon _{\text{py}} \epsilon _s\right)},\\
    \lambda_y &= -\frac{2 i \xi _p \xi_{\text{sy}}^2 V_y}{\epsilon _{\text{py}}^2 \epsilon _s} + \frac{2 \xi _p \xi _{\text{sx}} \xi _{\text{sy}} V_x}{\epsilon _{\text{px}} \left(\xi_{\text{sy}}^2-\epsilon _{\text{py}} \epsilon _s\right)}.
    \label{RashbaTerms}
\end{split}
\end{equation}
Note that the first-order contribution in Stark interaction on $\lambda_x$ magnitude depends on the difference of the electric dipoles at two consecutive sites $\imath$ and $\jmath$, so even though this is the term of the highest order, it is not necessarily the largest in magnitude, therefore, we consider that the second order terms are important for this description. In fact, the estimated values for the largest contributions are $\lambda_x\sim0.15$ meV and $\lambda_y\sim 1.2$ meV (see appendix B),  where we have considered that the angle of inclination of the hydrogen bonds with respect to the helix axis is very small, so $\xi_{sx} $ is negligible against $\xi_{sy}$.

The full SO effective interaction can be written as, $\mathcal{H}_{SO} = \mathcal{H}_{so} + \mathcal{H}_{R}$. The properties of the system will be determined mainly by the lowest order terms of (\ref{Heffective}). However, in case of mechanical deformations, higher order terms may be are relevant, so we consider here interactions up to second order in $\xi_{sy}$ and first order in $\xi_{sx}$. Then, the spin interactions of the effective Hamiltonian are determined mostly by the intrinsic SO, and the Rashba contribution become of comparable size in the case of mechanical deformations.

%****************************************************************************************************************************
\section{Bloch Space Hamiltonian}\label{sectionIII}
Consider a local cartesian coordinate system that is on top of an atom, then each atom on the chain will have the same system. The nearest neighbor atoms are described by the following vectors in the local system,
%{\color{red} It is this coordinate system the same Global coordinate system??}:
\begin{equation}
    \mathbf{\tau}^\pm = \frac{r}{\sqrt{2}}(\mathbf{e}_z\pm\mathbf{e}_x)\pm\frac{h}{4}\mathbf{e}_y.
\end{equation}
Considering only first nearest neighbors interaction, the Hamiltonian can be taken as the Bloch sum of matrix elements. Considering $k_z = 0$ and assuming that the contribution of each site is independent with nearest neighbor interaction only, the Bloch expansion can be obtained as,
\begin{equation}
\begin{split}\label{BlockExp}
    \mathcal{H}(k) &= \frac{1}{N} \sum_{\imath=1}^N \sum_{\jmath=1}^N e^{i \mathbf{k}\cdot\mathbf{R}_{\imath \jmath}}\langle \phi_{\imath}|\mathcal{H}|\phi_{\jmath} \rangle\\
    &= \frac{1}{N} \sum_{\imath=1}^N \left( \sum_{\jmath = \imath}  \langle \phi_{\imath}|\mathcal{H}|\phi_{\imath} \rangle +  \sum_{\jmath \neq \imath} e^{i \mathbf{k}\cdot\mathbf{R}_{\imath \jmath}}\langle \phi_{\imath}|\mathcal{H}|\phi_{\jmath} \rangle\right)\\
    &= \frac{1}{N} \sum_{\imath=1}^N  (\epsilon_\pi \mathbf{1}_s + V_z  f(k) \mathbf{1}_s + g(k)((\bm{\alpha} + \bm{\lambda}) \times \mathbf{s})_z\\
    &= \epsilon_\pi \mathbf{1}_s + V_z  f(k) \mathbf{1}_s + g(k)((\bm{\alpha} + \bm{\lambda}) \times \mathbf{s})_z.
\end{split}
\end{equation}
where we have only taken nearest neighbor couplings and strict periodicity of the lattice turn by turn. In Eq.\ref{BlockExp} $\phi_{\imath}$ are the orbitals per unit cell and $N$ is the number of the unit cells in the molecule. 
This model considers an approximate structure, shown in Fig. \ref{fig:structure}a, where the angle, $\Delta\phi$, between successive bases is smaller than the angle for real oligopeptides\cite{Pauling1951}. The latter assumption is not quite correct for oligopeptides since there is a small incommensurability (non-periodicity in the axial direction) of the potential when one goes from one turn to the next. This is an approximation of the model
%{\color{magenta}Here actually we have to approximation: first the elements we are ignoring in the diagonal part related to Stark correction to the Pz side energies, and next assuming some periodicity although it is not quite like that, but a convenient approximation, producing a slight increase in spin polarization in comparison to the real value with non periodic axial direction. I think both approximation should be mention, citing the IV section, showing the atomic SO values would not be to different to the real value. About the stark correction, mentioning that it is not a SO route and can be neglected in this framework (restoring inversion symmetry?)}.

The helix can be considered as a one dimensional system in the local frame that satisfies $\eta = \tan(h/r)$. Then, the one dimensional $k$ vector is proportional to $r'=r/\sqrt{2} + \eta h /4$, and the $k$ functions are,
\begin{equation}\label{k_funcs}
    f(k) = \cos\left(  r' k\right), \quad g(k) = \sin\left(  r' k \right) .
\end{equation}
The spectra of the system can be obtained by solving the secular equation
\begin{equation}\label{secular}
    \det(\mathcal{H}(k)-E \mathbf{S})=0,
\end{equation}
where $\mathbf{S}$ is the overlap matrix and we assuming that the eigen functions are orthogonals, such that $S=1$.  By solving the full system (\ref{secular}) we obtain the spectra of the system for the two spin species, and is given by,
\begin{equation}\label{fulldisperssion}
    E_\pm (k) = V_z \cos(r' k ) \pm \abs{\bm{\alpha}+\bm{\lambda}} \sin(r' k),
\end{equation}
where each band correspond to a different spin species.
%\begin{figure}[h!]
%  	  \centering
% 	  \includegraphics[width=0.45\textwidth]{kineticband.eps}
%  	  \caption{Kinetic spin degenerate bands with $K_\pm$ points(black) and a doped 
%         point (orange) taken at $k_F=3\pi/5$. The values for the parameters are 
% described in %Table \ref{tab:parameters} and Table \ref{tab:effective}.}
%  	  \label{fig:kineticBands}
%\end{figure}
\subsection{Hamiltonian in vicinity of half filling}
When the molecule is freestanding, delocalized electrons of $\pi$ space can be considered to be half filled. Consider that the Fermi energy of $p_z$ orbital is $\epsilon_F=0$, and spin interactions are perturbations with respect to kinetic energy. By solving (\ref{secular}) only for the kinetic component at half filling, $ \epsilon_F = V_z\cos( k_F) = 0$, then, the Fermi vector is $k_F = \pi/2$. To describe the physics in the vicinity of the Fermi level, let us consider a small perturbation $q$ around $k_F$, such that $k = k_F - \mathbf{q}$, and $0<|\mathbf{q}|<<1$. Then, the Bloch expansion of the system, (\ref{BlockExp}) can be approximated as,
\begin{equation}
    \mathcal{H}_{1/2}(q) = \epsilon_\pi + V_z q + ((\bm{\alpha} + \bm{\lambda}) \times \mathbf{s})_z.
\end{equation}
The spectra of the system %(Fig.\ref{fig:disperssion})% 
shows that the bands do not cross each other, they are always separated by a constant gap between spin up and spin down states of the order of $|\alpha|\sim 10^{-2}$eV. In such a system, the SO interaction is not coupled to momentum in the vicinity of $K_\pm$. 
%\begin{figure}[h!]
%  	  \centering
%  	  \includegraphics[width=0.45\textwidth]{disperssion.eps}
% 	  \caption{Dispersion around the $K_-$ point for small momenta for (red curve) %spin down, and (blue) spin up states.}
%  	  \label{fig:disperssion}
%\end{figure}
%\subsection{Doped Continuum Model}
Nevertheless, molecular contact with an environment, either a surface or surrounding structure will dope the system due to difference in electro-negativity. We must then consider an energy shift by above or below $\epsilon_F=0$. One can expand (\ref{Heffective}) around the doped energy, and the resulting expression has a spin component linear in momentum. Let us consider a small deviation from $k_F$, that is, $k'=3\pi/5$. The effective Hamiltonian around $k'$ is,
\begin{multline}
    \mathcal{H}_{k'} (q) = \epsilon_\pi + V_z \left( \frac{1-\sqrt{5}}{4}-\sqrt{\frac{5+\sqrt{5}}{8}}q \right)\\ + ((\bm{\alpha} + \bm{\lambda}) \times \mathbf{s})_z \left( \frac{1-\sqrt{5}}{4}q+\sqrt{\frac{5+\sqrt{5}}{8}} \right).
\end{multline}
Coupling between momentum and spin causes wavefunctions with a chiral component that increases approaching a crossing point at $k=0$. 

The previous Hamiltonian, aside from the geometrical details that determine the SO strength to within tens of meV, has the same form as that of DNA\cite{Varela2016} and of helicene\cite{HeliceneMujica} and leads to polarized electron transport, as has been reported experimentally\cite{Aragones,Kiran}.
%\section{Computational Methods}
%Density 

\section{Spin active deformation potentials}\label{sectionIV}
In this section we show the behavior, under mechanical deformations, of the SOC magnitudes. The response to deformations depends on the geometrical relations of the orbitals involved and will serve to provide an experimental probe to the model\cite{Kiran}. Although DNA and Oligopeptides are helices, the orbitals involved are quite different and thus should be distinguishable in a mechanical probe.

\begin{figure}[h!]
  	  \centering
  	  \includegraphics[width=0.3\textwidth]{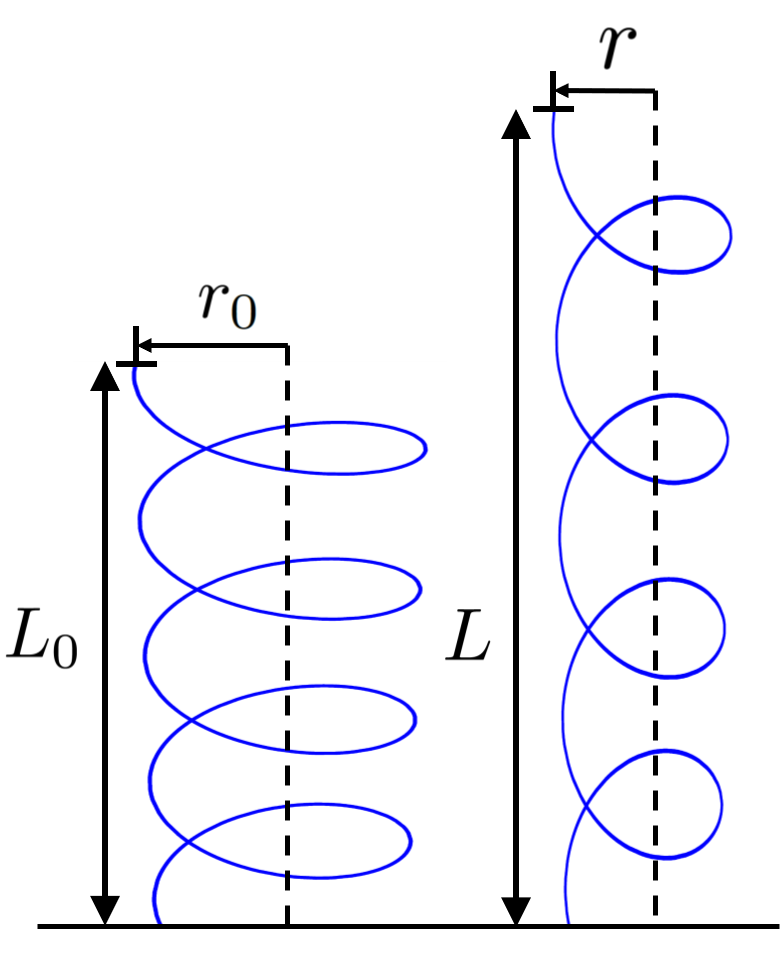}
  	  \caption{Graphical representation of a mechanical deformation setup. Left: Oligopeptide in initial structure $r_0$ and $L_0$. Right: Stretched structure along the helical axis to $r$ and $L$.}
  	  \label{fig:Helix}
\end{figure}

We consider stretching and/or compressing of the oligopeptide model in the form shown in the schematic Fig. \ref{fig:Helix}. 
In the deformation scheme, we consider that the rotation angle $\Delta\varphi$ (see Fig.\ref{fig:structure}) between consecutive atoms does not change for small deformations. The longitudinal strain is defined as $\varepsilon=(L-L_0)/L_0$ where $L_0$ and $L$ are the initial and final lengths of the helix, respectively. A change in $\varepsilon$ implies a change on the radius and pitch, such that $r=r_0(1-\nu \varepsilon)$ and $h=h_0(1+\varepsilon)$, where $\nu$ is the Poisson ratio of the helix\cite{OligopeptideMechanic,OligopeptideMatrix}. The deformation changes the relative distances between orbitals, so the magnitude of the vector connecting two neighboring sites is written in the form
\begin{equation}
    R_{\jmath\imath}(\varepsilon)=\sqrt{r_{0}^2 (1-\nu\varepsilon)^2 +h_0^2(1+\varepsilon)^2/16}.
\end{equation}

The expressions for the SO intrinsic terms are
\begin{equation}
\alpha_x=\frac{2\hbar^2\xi_p}{m \epsilon_p (R_{\jmath\imath}(\varepsilon))^2}\left(\kappa_{pp}^{\pi}- \frac{r_0^2 (1-\nu\varepsilon)^2(\kappa_{pp}^{\sigma}-\kappa_{pp}^{\pi})}{(R_{\jmath\imath}(\varepsilon))^2}\right),    
\end{equation}
and 
\begin{equation}
\alpha_{y}=-\frac{2\hbar^2\xi_p r_0 h_0(1-\nu\varepsilon)(1+\varepsilon)(\kappa_{pp}^{\sigma}-\kappa_{pp}^{\pi})}{m\epsilon_p (R_{\jmath\imath}(\varepsilon))^4},    
\end{equation}
where we have considered that $\epsilon_{p}^{\pi}=\epsilon_{p}^{\sigma}=\epsilon_p$. For the first order dependence on $\varepsilon$ we have:
\begin{equation}
\alpha_x\approx\alpha_x^{(\varepsilon=0)} - 16r_0\xi_pC\nu(\kappa_{pp}^{\sigma} - \kappa_{pp}^{\pi})  \varepsilon+ \ldots,  
\end{equation}
and 
\begin{equation}
\alpha_y\approx\alpha_y^{(\varepsilon=0)}+8h_0\xi_pC(1-\nu)(\kappa_{pp}^{\sigma} - \kappa_{pp}^{\pi})\varepsilon+\ldots,
\end{equation}
where we have defined the constant $$C=\frac{64\hbar^2 r_0}{m\epsilon_p(h_0^2+16r_0^2)^2}$$.
\begin{figure}[H]
  	  \centering
  	  \includegraphics[width=8.5cm]{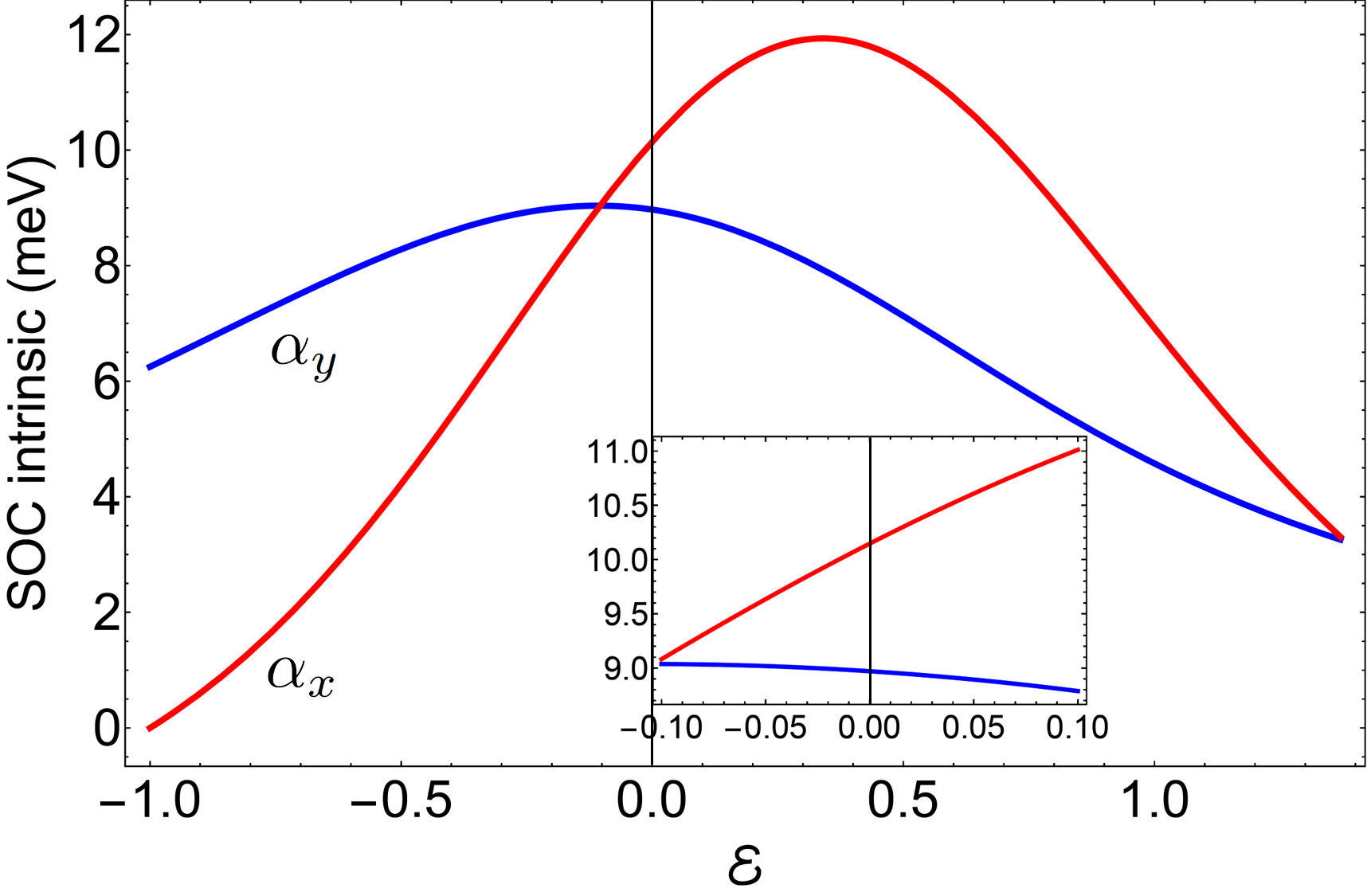}
  	  \caption{SOC intrinsic intensities $\alpha_x$ and $\alpha_y$ versus deformation $\varepsilon$. We used $r_0=0.23$ nm, $h_0=0.54$ nm, $\Delta\varphi=\pi/2$ and $\nu=0.5$. For $\varepsilon=0$, the intensity of the interactions are $\alpha_x=8.97$ meV and $\alpha_y=10.20$ meV.}
  	  \label{Intrinsic}
\end{figure}
The coefficients of the linear in $\varepsilon$ are the spin-dependent {\it deformation potentials}\cite{book:winkler} for the intrinsic interaction. 

Figure \ref{Intrinsic} displays the intrinsic SOC magnitudes as a function of the deformation $\varepsilon$. Positive values for $\varepsilon$ show the behavior when the helix is stretched and negative values when it is compressed. For small values of deformations, $\alpha_x$ grows with a stretch at the same time as $ \alpha_y $ decreases (see inset in Fig.\ref{Intrinsic}). This different behavior is due to the initial relative orientation of the orbitals. However, the longitudinal deformation that arises from considering the SO net magnitude, has an increase during stretching and a decrease when compressed, the same behavior of the corresponding deformation configurations of the SO obtained for the DNA\cite{VarelaJCP}. This behavior has a maximum that represents the optimum strain value for maximum SOC, in this case up to $20$ meV, for a deformation of $20\%$ with respect to the initial length, that is, the magnitude of the interaction doubles with respect to the value without deformation.  Nevertheless, this stretch may alter the assumed structure as hydrogen bonding may rupture\cite{OligopeptideMechanic}. We have taken the Slater-Koster elements as decreasing with the square of the distance (for $\varepsilon>0$) and/or the orbitals become orthogonal (for $\varepsilon<0$)\cite{book:Harrison}.

The expressions for the Rashba terms as a function of deformation are 
\begin{equation}
\lambda_x=\frac{\hbar^2\xi_p \kappa_{sp}^{\sigma}r_0(1-\nu\varepsilon)(\xi_{sy,\imath}(\varepsilon)-\xi_{sy,\jmath}(\varepsilon))}{m\epsilon_{p}\epsilon_s(R_{\jmath\imath}(\varepsilon))^3},
\end{equation}
and 
\begin{equation}
\lambda_y=\frac{2\hbar^2\xi_p(\xi_{sy}(\varepsilon))^2r_0 h_0 (1-\nu\varepsilon)(1+\varepsilon)(\kappa_{pp}^{\sigma}-\kappa_{pp}^{\pi}) }{m\epsilon_s\epsilon_{p}^2(R_{\jmath\imath}(\varepsilon))^4},
\end{equation}  
where we only consider the first terms in equation \ref{RashbaTerms} for $\lambda_x$ and $\lambda_y$, since they are the most significant in magnitude. The Stark parameters will be modulated by the change in the hydrogen bond polarization due to the longitudinal deformation in the same form that is in Ref\cite{VarelaJCP}.

\begin{figure}[h!]
 \centering
  \includegraphics[width=8.5cm]{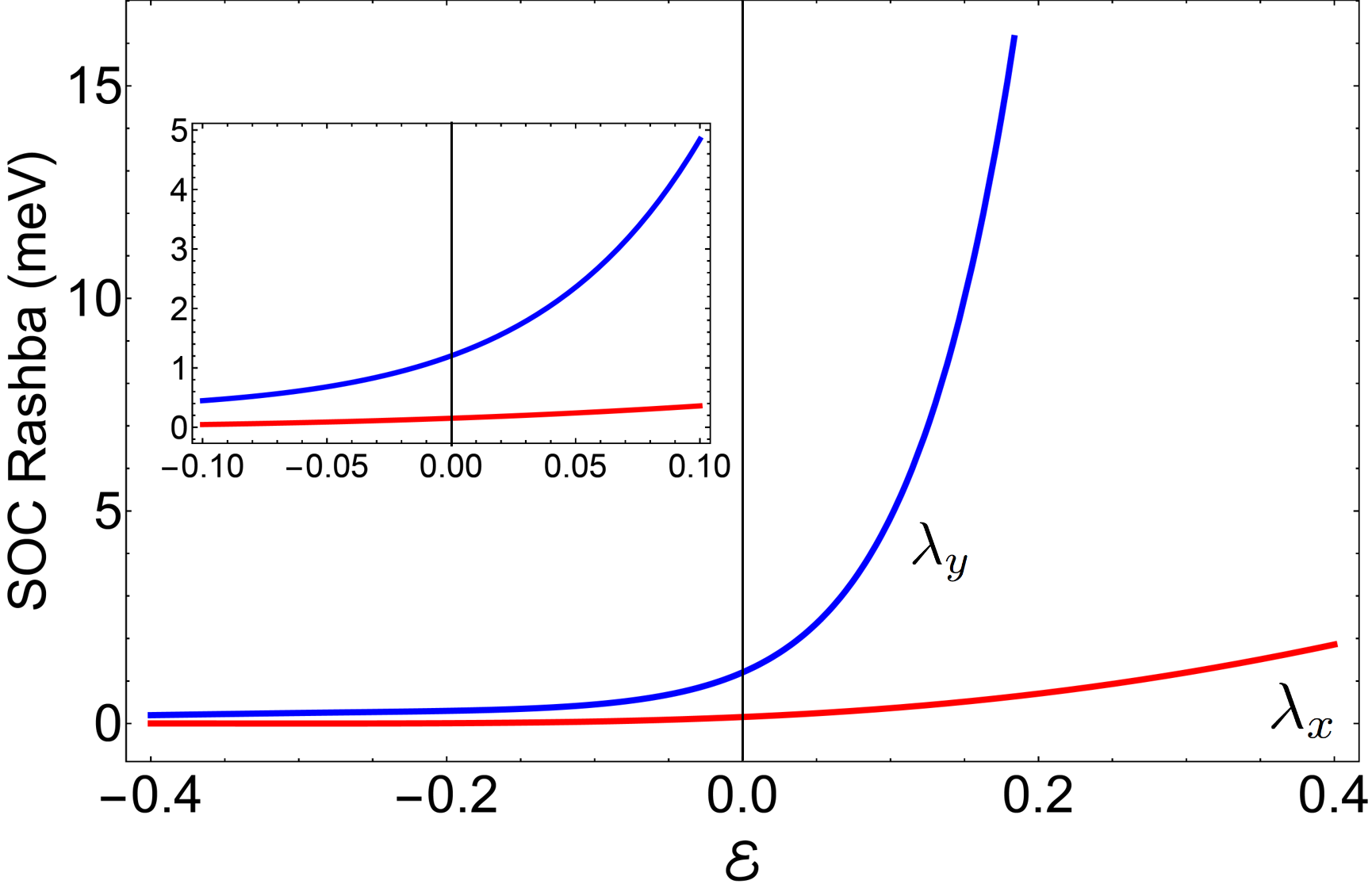}
 \caption{Rashba magnitudes $\lambda_x$ and $\lambda_y$ versus deformation $\varepsilon$. We used $r_0=0.23$ nm, $h_0=0.54$ nm, $\Delta\varphi=\pi/2$ and $\nu=0.5$. For $\varepsilon=0$, the intensity of the interactions are $\lambda_x=0.15$ meV and $\lambda_y=1.2$ meV. Streching the helix ($\varepsilon>0$) increases the Rashba coupling while compressing decreases it.}
 \label{Rashbadefor}
\end{figure}

The Rashba terms are proportional to the electric fields of the dipoles, therefore, when stretching the helix the relative distances between the orbitals become large, which decreases the Slater-Koster elements, but the length of the dipoles increase and this behavior is dominant such that it increases the Rashba magnitude, as it is shown in Fig. \ref{Rashbadefor}. This is the opposite behavior seen for DNA\cite{VarelaJCP}.

For the first order dependence of the Rashba interaction on $\varepsilon$ we have:
\begin{equation}
\lambda_{x}\approx\lambda_{x}^{(\varepsilon=0)}+\frac{\kappa_{sp}\xi_pC\nu (\xi_{sy,\imath} -\xi_{sy,\jmath})}{\epsilon_s(h_0^2+16r_0^2)^{-1/2}}\varepsilon + \ldots,
\end{equation}
and 
\begin{equation}
\lambda_{y}\approx\lambda_{y}^{(\varepsilon=0)}+   \frac{8h_0\xi_pC(1-\nu)(\xi_{sp})^2(\kappa_{pp}^{\pi} - \kappa_{pp}^{\sigma})}{\epsilon_s\epsilon_p}\varepsilon+\ldots,
\end{equation}
where the linear in $\varepsilon$ terms are the spin-dependent deformation potentials of the Rashba coupling. Note that $\lambda_x$ is sensitive to differences in the Stark interaction at two different sites. On the other hand, $\lambda_y$ depends on the square of the Stark interaction. Although these features may lead to a smaller size of the SOC they are actually enhanced by deformation to be comparable to the intrinsic contribution (see Fig.\ref{Rashbadefor}).

In the deformation range of $10\%$, the magnitude of the Rashba interaction can increase up to 5 times its initial value (inset, Fig.\ref{Rashbadefor}). This result is opposite to the corresponding deformation previously obtained in the DNA, where stretching the helix longitudinally, decreased the polarization of the hydrogen bonds that in that case were oriented transversely to deformation. 

The behavior under deformation agrees qualitatively with that found in experiments\cite{Kiran}, where spin polarization decreases with the compression of oligopeptides under an applied force. It is important to note that the quadratic terms in $\lambda_y$ are much more sensitive to deformation than the first order term ($\lambda_x$), so deformations during experimental tests can induce higher order terms in interactions to contribute significantly to the magnitude of the effective coupling.

\section{Summary and Conclusions}\label{conclusions}
In this work we have studied the nature of spin interactions of oligopeptides including the effects of internal electric fields and SOC. We built a minimal analytic tight-binding model to describe the mobile electrons of the system in a helical geometry using the Slater-Koster approach. We assumed mobile electron spring from carboxyl group double bonds attached to Amine groups through hydrogen bonding. Perturbative band folding then yields effective SO interactions of the Intrinsic and Rashba types. We find a rich interplay between intrinsic and Rashba SOC's that allows manipulation of the spin polarization of oligopeptides under mechanical longitudinal deformation probes. The low-energy effective Hamiltonian in the vicinity of the half filling Fermi level shows the same form of Hamiltonians derived for DNA and Helicene that have shown spin-polarization, explaining features of the CISS effect. The response to deformations expressed as spin-dependent deformation potentials are consistent with the results of ref.\onlinecite{Kiran} and opposite trends to the results found for DNA. These results both make strong predictions to verify our orbital model and open possibility of mechanical probes to spintronic properties of biological molecules.

\acknowledgements{This work was supported by CEPRA VIII Grant XII-2108-06 ``Mechanical Spectroscopy".}

%****************************************************************************************************************************
\appendix

\section*{Appendix A: Slater-Koster integrals}\label{sec:SK}
The overlap $E_{\mu\mu'}^{\imath\jmath}$ between orbitals $\mu$ and $\mu'$ that correspond to the site $\imath$ and $\jmath$ respectively, can be obtained using the expression \cite{Varela2016,Geyer2019} 
\begin{multline}
	E_{\mu \mu'}^{\imath\jmath} = \langle \mu_{\imath} | V | \mu_{\jmath}' \rangle = (\mathbf{n}(\mu_{\imath}),\mathbf{n}(\mu_{\jmath}'))V_{\mu \mu'}^\pi \\+ \frac{(\mathbf{n}(\mu_{\imath}),\mathbf{R_{\jmath\imath}}) (\mathbf{n}(\mu_{\jmath}'),\mathbf{R_{\jmath\imath}})}{(\mathbf{R_{\jmath\imath}},\mathbf{R_{\jmath\imath}})} (V_{\mu \mu'}^\sigma-V_{\mu \mu'}^\pi),
\label{SK}
\end{multline}
where $\mathbf{n}(\mu_\jmath)$ is the unit vector on the direction of the orbital $\mu$ of site $\jmath$, $\mathbf{R_{\jmath\imath}}$ is the vector that connect two consecutive sites, and $V_{\mu \mu'}^\sigma$ and $V_{\mu \mu'}^\pi$ represent the Slater-Koster overlaps of the orbitals. 

The unit vector of each orbital in a local coordinate system (xyz) on site $\imath$ is given by
\begin{equation}
\begin{split}
    \hat{\mathbf{n}}(s_\imath)&=\hat{\mathbf{R}}_{\jmath\imath},\\
    \hat{\mathbf{n}}(x_\imath)&=-\sin(\varphi_\imath) \mathbf{e}_x+\cos(\varphi_\imath)\mathbf{e}_z,\\
    \hat{\mathbf{n}}(y_\imath)&=\mathbf{e}_y,\\
    \hat{\mathbf{n}}(z_\imath)&=\cos(\varphi_\imath) \mathbf{e}_x+\sin(\varphi_\imath)\mathbf{e}_z,
\end{split}
\end{equation}
The Slater-Koster terms have a dependence on the distance representing in the empirical expression in the literature\cite{book:Harrison},
\begin{equation}\label{Harrison}
    V_{\mu\mu'}^{\pi,\sigma}= \kappa_{\mu\mu'}^{\pi,\sigma} \frac{\hbar^2}{m R_{\jmath\imath}^2},
\end{equation}
where $m$ is the mass of the electron and $\kappa_{\mu\mu'}^{\pi,\sigma}$ depend on the specific set of orbitals or atoms. 

Without loss of generality we can assume that $E_{\mu\mu'}^{\imath \jmath}=0$, where $\mu = \{ s,p_x,p_y \}$, because those electrons form the bond. The Slater-Koster integrals that are relevant for transport processes, in terms of general parameters of the structure, are the following:
\begin{equation}
\begin{split}
    E_{zz}^{\imath\jmath}&=\langle z_\imath | V | z_\jmath\rangle = \\ &  \cos[\Delta\varphi] V_{pp}^\pi
    - \frac{r^2}{|\mathbf{R_{\jmath\imath}}|^2} (1-\cos[\Delta\varphi])^2 (V_{pp}^\sigma-V_{pp}^\pi)\\
    E_{zx}^{\imath\jmath}&=\langle z_\imath | V | x_\jmath\rangle = \\ & \sin[\Delta\varphi] \left( V_{pp}^\pi - \frac{r^2}{|\mathbf{R_{\jmath\imath}}|^2} (1-\cos[\Delta\varphi])(V_{pp}^\sigma-V_{pp}^\pi) \right)\\
    E_{zy}^{\imath\jmath}&=\langle z_\imath | V | y_\jmath\rangle =\\& - \frac{h r}{|\mathbf{R_{\jmath\imath}}|^2} (1-\cos[\Delta\varphi])(\jmath-\imath) (V_{pp}^\sigma-V_{pp}^\pi)\\
    E_{zs}^{\imath\jmath}&=\langle z_\imath|V|s_\jmath\rangle = \frac{r(1-\cos[\Delta\varphi])}{|\mathbf{R}_{\jmath\imath}|} V_{sp}^\sigma.
\end{split}
\end{equation}
Using the geometry shown in Fig. \ref{fig:structure}, i.e. $\Delta \phi= \pi/2$, the following symmetry relations are obtained:
\begin{equation}
\begin{split}
    V_{z}=E^{ij}_{zz} &= E^{ji}_{zz} = - \frac{r^2}{|\mathbf{R_{ji}}|^2} (V_{pp}^\sigma-V_{pp}^\pi) ,\\
    V_{s}=E^{ij}_{zs} &=E^{ji}_{zs} = E^{ij}_{sz} = E^{ji}_{sz} = \frac{r}{|\mathbf{R}_{ji}|} V_{sp}^\sigma,\\
    V_{x}=E^{ij}_{zx} &=-E^{ji}_{zx} =-E^{ij}_{xz} = E^{ji}_{xz}= V_{pp}^\pi - \frac{r^2}{|\mathbf{R_{ji}}|^2}(V_{pp}^\sigma-V_{pp}^\pi) ,\\
    V_{y}=E^{ij}_{zy} &=-E^{ji}_{zy} = -E^{ij}_{yz} = E^{ji}_{yz} =  - \frac{r h}{|\mathbf{R_{ji}}|^2} (V_{pp}^\sigma-V_{pp}^\pi).
\end{split}
\end{equation}
\section*{Appendix B: Parameters for the effective system}
We estimate the overlaps of the atomic wavefunctions using ref. \ref{Harrison}. The geometrical structure of the oligopeptide includes four atoms per turn and it does not differ significantly from realistic situations where oligopeptides are not strictly periodic from one turn to the next\cite{Pauling1951}. Atomic and structural parameters for the system are given in Table \ref{tab:parameters}. The SK and SO effective magnitudes are written in Table \ref{tab:effective}. 

\begin{table}[!htbp]
\caption{Left column: SK parameters for $s$ and $p$ orbitals from \cite{Varela2016}. Center column: Atomic parameters for carbon atoms from  \cite{Varela2016,Konschuh2010} Right column: Structural parameters used to describe the oligopeptide.}
\begin{center}
\begin{tabular}{|cc|cc|cc|}
\hline
 Parameter & eV & Parameter & eV & Parameter & \AA / rad.\\
\hline
$\kappa_{pp}^{\sigma} $ & -0.81 & $\epsilon_p$ & -8.97 & $r$ & 2.3 \\
$\kappa_{pp}^{\pi} $ & 3.24  & $\epsilon_{s}$ & -17.52 & $h$ & 5.4  \\
$\kappa_{sp} $ & 1.84& $\xi_p$ & 0.006& $\Delta \varphi$& $\pi/2$ \\
\hline
\end{tabular}\label{tab:parameters}
\end{center}
\end{table}

\begin{table}[!htbp]
\caption{Estimation of effective interactions for the system without deformations. Left column: Hopping interactions. Right column: SO interactions.}
\begin{center}
\begin{tabular}{|cc|cc|}
\hline
Parameter & eV & Parameter & meV\\
\hline
$V_s$ &3.786&$\alpha_x$ & 8.97\\
$V_x$ &-4.143&$\alpha_y$ & 10.20\\
$V_y$ &-7.666&$\lambda_x$ & 0.15\\
$V_z$ &-3.265&$\lambda_y$ & 1.2\\

\hline
\end{tabular}\label{tab:effective}
\end{center}
\end{table}

%*********************************************************************************************************************

%****************************************************************************************************************************

%\section*{Acknowledgment}
%
%We are gratefully to Dr. Xavier Garc\'ia for fruitfull discussion about the work and to provide
%clearness about the code of the simulation.

%----------------------------------------------------------------------------------------
%	REFERENCE LIST
%----------------------------------------------------------------------------------------
%\bibliographystyle{unsrt}
%\bibliography{biblio.bib}

\end{document}